\documentclass[twocolumn,showpacs,preprintnumbers,amsmath,amssymb,superscriptaddress]{revtex4}

\usepackage{graphicx}

\def\prtx{Pr$T$$_2$$X$$_{20}$}
\def\prtal{Pr$T$$_2$Al$_{20}$}

\def\prrhzn{PrRh$_2$Zn$_{20}$}
\def\prirzn{PrIr$_2$Zn$_{20}$}
\def\lairzn{LaIr$_2$Zn$_{20}$}

\def\prtial{PrTi$_2$Al$_{20}$}
\def\prval{PrV$_2$Al$_{20}$}

\def\tq{$T_{\rm Q}$}

\def\tc{$T_{\rm c}$}

\begin{document}




\title{Quadrupole-Driven Non-Fermi Liquid and Magnetic-Field Induced Heavy Fermion States in a Non-Kramers Doublet System}

\author{T. Onimaru}
\email{onimaru@hiroshima-u.ac.jp}
\affiliation{%
Department of Quantum Matter, Graduate School of Advanced Sciences of Matter, Hiroshima University, Higashi-Hiroshima 739-8530, Japan
}%
\author{K. Izawa}%
\affiliation{%
Department of Condensed Matter Physics, Graduate School of Science and Engineering, Tokyo Institute of Technology, Tokyo 152-8551, Japan
}%
\author{K. T. Matsumoto}%
\affiliation{%
Department of Quantum Matter, Graduate School of Advanced Sciences of Matter, Hiroshima University, Higashi-Hiroshima 739-8530, Japan
}%
\author{T. Yoshida}%
\affiliation{%
Department of Condensed Matter Physics, Graduate School of Science and Engineering, Tokyo Institute of Technology, Tokyo 152-8551, Japan
}%
\author{Y. Machida}%
\affiliation{%
Department of Condensed Matter Physics, Graduate School of Science and Engineering, Tokyo Institute of Technology, Tokyo 152-8551, Japan
}%
\author{T. Ikeura}%
\affiliation{%
Department of Condensed Matter Physics, Graduate School of Science and Engineering, Tokyo Institute of Technology, Tokyo 152-8551, Japan
}%
\author{K. Wakiya}%
\affiliation{%
Department of Quantum Matter, Graduate School of Advanced Sciences of Matter, Hiroshima University, Higashi-Hiroshima 739-8530, Japan
}%
\author{K. Umeo}%
\affiliation{%
Cryogenics and Instrumental analysis Division, N-BARD, Hiroshima University, Higashi-Hiroshima 739-8526, Japan
}%
\author{S. Kittaka}%
\affiliation{%
Institute for Solid State Physics, University of Tokyo, Kashiwa 277-8581, Japan
}%
\author{K. Araki}%
\affiliation{%
Institute for Solid State Physics, University of Tokyo, Kashiwa 277-8581, Japan
}%
\author{T. Sakakibara}%
\affiliation{%
Institute for Solid State Physics, University of Tokyo, Kashiwa 277-8581, Japan
}%
\author{T. Takabatake}%
\affiliation{%
Department of Quantum Matter, Graduate School of Advanced Sciences of Matter, Hiroshima University, Higashi-Hiroshima 739-8530, Japan
}%
\affiliation{%
Institute for Advanced Materials Research, Hiroshima University, Higashi-Hiroshima 739-8530, Japan
}%


\date{\today}

\begin{abstract}
Orbital degrees of freedom in condensed matters could play important roles in forming a variety of exotic electronic states by interacting with conduction electrons.
In 4$f$-electron systems, because of strong intra-atomic spin-orbit coupling, an orbitally degenerate state inherently carries quadrupolar degrees of freedom.
The present work has focussed on a purely quadrupole-active system {\prirzn} showing superconductivity in the presence of an antiferroquadrupole order at {\tq} = 0.11 K.
We observed non-Fermi liquid (NFL) behaviors emerging in the electrical resistivity ${\rho}$ and the 4$f$ contribution to the specific heat, $C_{4f}$, in the paramagnetic state at $T$ $>$ {\tq}.
Moreover, in magnetic fields $B$ ${\le}$ 6 T, all data set of ${\rho}(T)$ and $C_{4f}(T)$ are well scaled with characteristic temperatures $T_{0}$'s.
This is the first observation of the NFL state in the nonmagnetic quadrupole-active system, whose origin is intrinsically different from that observed in the vicinity of the conventional quantum critical point. It implies possible formation of a quadrupole Kondo lattice resulting from hybridization between the quadrupoles and the conduction electrons with an energy scale of $k_{\rm B}{T}_{0}$. 
At ${T}{\le}$0.13 K, ${\rho}(T)$ and $C_{4f}(T)$ exhibit anomalies as $B$ approaches 5 T.
This is the manifestation of a field-induced crossover toward a Fermi-liquid ground state in the quadrupole Kondo lattice.
\end{abstract}

\maketitle

\section{Introduction}

In metals and alloys, localized $d$ and/or $f$ electrons inherently possessing `spin' and `orbital' degrees of freedom are dominant sources of not only various magnetic phenomena but also unconventional superconductivity.
Extensive studies on a lot of `spin'-active systems have successfully revealed the variety of the magnetic phenomena arising from competition of the inter-site Ruderman-Kittel-Kasuya-Yosida (RKKY) interaction and the on-site magnetic Kondo effect, which was represented as the well-known `Doniach diagram'.
On the other hand, there is less variety of researches on `orbital'-driven physics. Although the `orbital' itself must have potential variety, there are few evidences of interplay between the orbital degrees of freedom and the conduction electron.

In the $f$-electron systems, strong intra-atomic spin-orbit coupling forces the spin and orbital degrees of freedom to be described in terms of the total angular momentum \textit{\textbf J}.
When the magnetic moment due to \textit{\textbf J}
interacts strongly with itinerant conduction electrons, physically observable quantities follow the Fermi-liquid (FL) model of Landau, 
which is known as heavy fermion state.
When some conditions were fulfilled unexpectedly, the FL state could become unstable, instead, an anomalous metallic state would emerge, so-called non-Fermi liquid (NFL) state.\cite{Lohneysen,Umeo,Si01,Custers03}
In the vicinity of the quantum critical point, there emerges unconventional superconductivity.\cite{Mathur98}

On the other hand, emergence of a different type of NFL state was predicted theoretically, in case electric quadrupoles of the localized $f$ electrons, which become active only in an orbitally degenerate state, interact with the conduction electrons, that is the quadrupole (two-channel) Kondo effect.\cite{Cox98}
It is quite different from the ordinary (single-channel) $magnetic$ Kondo effect in terms of scattering process of the conduction electrons; the scattering source is not magnetic dipole but the time-reversal electric quadrupole. 
Thereby, the impurity quadrupoles were overcompensated by the conduction electrons, leading to the residual entropy of (0.5)$R$ln2 and the NFL behavior:
the magnetic susceptibility $\chi$ and the specific heat divided by temperature, ${C}{/}{T}$, show ln$T$ dependence, and the electrical resistivity follows ${\rho}-{\rho}_{0}{\propto}1{+}{A}\sqrt{T}$, where ${\rho}_{0}$ is residual resistivity and $A$ is a coefficient.\cite{Affleck91,Cox98}
NFL state and the residual entropy were also expected in a quadrupole Kondo `lattice', in which quadrupole moments are periodically placed.\cite{Jarrell96}
Very recently, the peculiar temperature dependence of ${\rho}(T)$ and ${C}(T)$ have been theoretically predicted.\cite{Tsuruta99,Tsuruta15}
Moreover, in the lattice model, new types of electronic ordered states have been proposed.\cite{Hoshino11,Hoshino13,Hastatic}
Nevertheless, there is no experimental evidence on the exotic states, probably because there are rare systems carrying purely active quadrupoles.

Over the past few decades, experimental efforts have been paid to address the issues that could arise from the impurity quadrupole Kondo effect.\cite{Stewart01}
NFL behaviors were observed in uranium- and praseodymium-based systems with 5$f^2$ and 4$f^2$ configurations, respectively, such as U$_{x}$Y$_{1-x}$Pd$_{3}$\cite{Seaman91,McEwen03}, UBe$_{13}$\cite{McElfresh93}, U$_{x}$Th$_{1-x}$Be$_{13}$\cite{Aliev93}, U$_{x}$Th$_{1-x}$Ru$_2$Si$_2$\cite{Amitsuka94}, and Pr$_{x}$La$_{1-x}$Pb$_3$\cite{Kawae06}. 
However, there is no firm experimental evidence for the impurity quadrupole Kondo effect, since atomic disorder or uncertainty in the crystalline electric field (CEF) ground states particularly in the U-based systems prevent the clarification.

Recently, a family of the praseodymium-based systems {\prtx} ($T$: transition metals, $X$: Al, Zn, and Cd) have emerged as a prototype to study the quadrupole Kondo effect, since the CEF ground state is a non-Kramers doubly degenerated state, that is labeled as the ${\Gamma}_3$ doublet in the cubic $T_{\rm d}$ point group, having no magnetic dipoles but electric quadrupoles.\cite{Nasch97,Onimaru11,Sakai11,Sato12,Iwasa13,Yazici15} 
In {\prirzn}, an AFQ order occurs at {\tq} = 0.11 K.\cite{Onimaru11}
Although the entropy release of $R$ln2 is expected from an order of the doubly degenerated CEF ground state, the entropy at {\tq} estimated from the specific heat is only 20\% of $R$ln2.
Therefore, there should be another mechanism, except  the AFQ order, which consumes the rest of the entropy above {\tq}.
On the other hand, below {\tq}, a superconducting transition sets in at {\tc} = 0.05 K, suggesting a possible interplay between the quadrupole fluctuations and the superconducting Cooper-pair formation.\cite{Onimaru10, Onimaru11, Ishii11, Higemoto12}
The coexistence of superconductivity with quadrupole order also manifests in isostructural compounds {\prrhzn}, {\prtial}, and {\prval}.\cite{Onimaru12, Sakai12, Tsujimoto14,Matsubayashi12}
In {\prtial} and {\prval}, the strong hybridization between the 4$f$ elections and the conduction electrons was revealed by the Al-NMR and Pr 3$d$-4$f$ resonant photoemission measurements.\cite{Tokunaga13,Matsunami11}
The large Seebeck coefficient divided by temperature ${S}{/}{T}$ observed in {\prtal} ($T$ = Ti, V and Ta) also suggests the strong hybridization effect, although that for {\prirzn} at $B=0$ is two or three orders smaller than those of {\prtal}.\cite{Machida15}
{\prval} exhibits NFL behavior above {\tq} = 0.6 K; the 4$f$ contributions in both ${\rho}$ and ${\chi}$ follow $\sqrt{T}$ for 2${<}{T}{<}$20 K\cite{Sakai11}, however, in the temperature range, the magnetic degrees of freedom in the first excited state must interfere in the NFL behavior.
Quadrupolar quantum criticality induced by application of magnetic field has been proposed.\cite{Shimura15}

In the present work, we study the transport, thermodynamic and magnetic properties for the non-Kramers doublet system {\prirzn} in magnetic fields applied along the [100] direction.
For \textit{\textbf B} $||$ [100], the AFQ order collapses at around 5 T, whose value is much lower than the critical fields for the AFQ order in the [110] and [111] field directions.\cite{Onimaru11,Ishii11,Hattori14} 
Since high quality of the sample is necessary to reveal the inherent behavior arising from the quadrupolar degrees of freedom as described above,
we have chosen single-cryslltaine samples with low residual resistivity of 0.2 ${\mu}{\Omega}$ cm (residual resistivity ratio, RRR$>$100).\cite{Onimaru11} 
The manifestation of the AFQ order guarantees the high-quality of the crystals, otherwise it is destroyed by the atomic disorder.
Our measurements of the specific heat and the electrical resistivity in magnetic fields have revealed emergence of the NFL state possibly arising from formation of a quadrupole Kondo lattice. 
Furthermore, we have observed a magnetic-field induced crossover from the NFL state to an exotic heavy-fermion ground state in the quadrupole Kondo lattice.

\section{Experimental}

Single-crystalline samples of {\prirzn} used in the present work were grown by using the melt-growth method described in the previous papers.\cite{Onimaru10}
Magnetization was measured using a commercial SQUID magnetometer (Quantum Design MPMS) between 1.9 and 350 K in magnetic fields up to 5 T. Magnetization measurements at low temperatures down to 0.045 K were performed by a capacitive Faraday method using a high-resolution capacitive force-sensing device installed in a $^3$He-$^4$He dilution refrigerator.\cite{Sakakibara94}
Electrical resistance was measured by a standard four-probe dc method in a laboratory-built system with a $^3$He-$^4$He dilution refrigerator. Thermopower was measured using a laboratory-made probe by applying a temperature difference of 0.04-0.3 K along a bar-shaped sample.
Specific heat in magnetic fields was measured by a relaxation method at temperatures between 0.4 K and 300 K. 
The measurements at low temperatures down to 0.06 K were done under quasi-adiabatic conditions with a $^3$He-$^4$He dilution refrigerator equipped with a superconducting magnet of 12 T.

\section{Results and discussion}


\begin{figure}[t]
\includegraphics[scale=0.5]{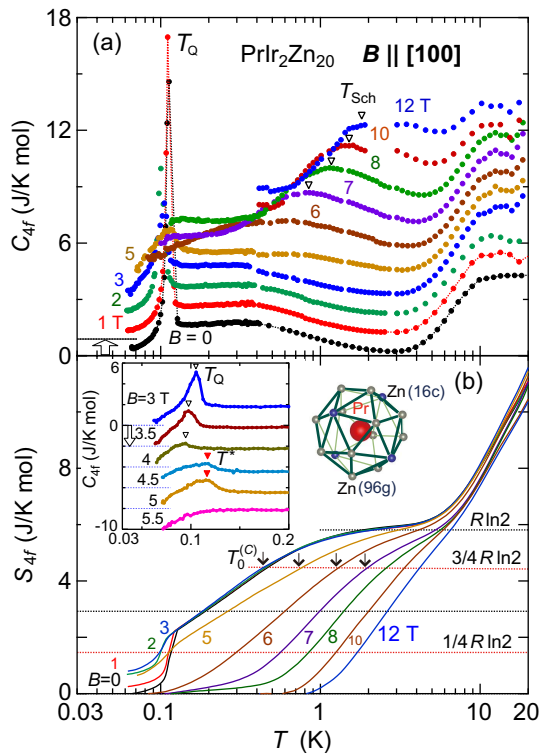}
\caption{(color online) (a) 4$f$ contribution to the specific heat $C_{4f}$ of {\prirzn} in magnetic fields up to 12 T applied along the [100] direction.
The data of $C_{4f}$ are vertically offset for clarity. {\tq} indicates the temperature of AFQ order. The open triangles are $T_{\rm Sch}$, where Schottky-type peaks appear as a result of the splitting of the ground state doublet by applying the magnetic fields.
(b) Entropy $S_{4f}$ estimated by integrating $C_{4f}(T){/}{T}$ in various fields.The arrow indicates characteristic temperature ${T}_{0}^{(C)}$ defined as the temperature where $S_{4f}$ reaches $\frac{3}{4}$$R$ln2.
The inset shows the data of $C_{4f}$ for 3$\le$$B$$\le$5.5 T plotted with vertical offsets. The open and closed triangles, respectively, represent {\tq} and a crossover temperature $T^{*}$ emerging only in ${B}{=}$4.5 and 5 T, which will be described in the text.
A Pr ion encapsulated into the symmetric Zn-cage, which is formed by four Zn atoms at the 16$c$ site and twelve Zn atoms at the 96$g$ site, is also shown in the inset.
}
\label{HC}
\end{figure}

\subsection{Specific heat}

Figure \ref{HC} (a) shows the temperature dependence of the 4$f$ contribution to the specific heat, $C_{4f}$, of {\prirzn} in magnetic fields ${B}$ ${\le}$ 12 T applied along the [100] direction. 
To estimate the 4$f$ contribution $C_{4f}$ to the total specific heat $C$, we subtracted the nuclear and phonon contributions $C_{\rm nuc}$ and $C_{\rm ph}$ as described below.
The hamiltonian of a nuclear spin of a Pr nucleus, ${I}{=}$5/2, in magnetic field \textit{\textbf B} can be represented as
\begin{equation}
\mathcal{H}_{\rm nuc} = A_{\rm hf} \textit{\textbf J} \cdot \textit{\textbf B} - g_{\rm N} {\mu}_{\rm N} \textit{\textbf B} \cdot \textit{\textbf I},
\end{equation}
where $A_{\rm hf}{=}$0.052 K is a coupling constant of hyper-fine interaction of a Pr ion \cite{Kondo61}, and \textit{\textbf J}, $g_{\rm N}{=}$1.72, and ${\mu}_{\rm N}$ are the total angular momentum of 4$f$ electrons, nuclear $g$-factor, and nuclear magneton, respectively.
The component of \textit{\textbf J} along the magnetic field direction was regarded as the isothermal magnetization measured at ${T}{=}$0.045 K. 
Thereby, the nuclear specific heat was estimated by taking the eigenvalues of the hamiltonian into consideration. 
The nuclear contribution is zero in ${B}{=}$0. 
Applying magnetic field, it gradually increases on cooling, $e.g.$, in $B$ = 3 T,  ${C}_{\rm nuc}$ is increased up to about 3.5 J/K mol at 0.1 K.
The contribution of the phonon was subtracted by using the specific heat of the La analog {\lairzn}.
The main panel of Fig. \ref{HC} (b) shows the entropy $S_{4f}$ estimated by integrating $C_{4f}{/}{T}$.
The value of $S_{4f}$ reaches $R$ln2 at 2 K, supporting that the physical properties at $T$ $<$ 2 K are governed by the doublet ground state.

\begin{figure}[t]
\includegraphics[scale=0.5]{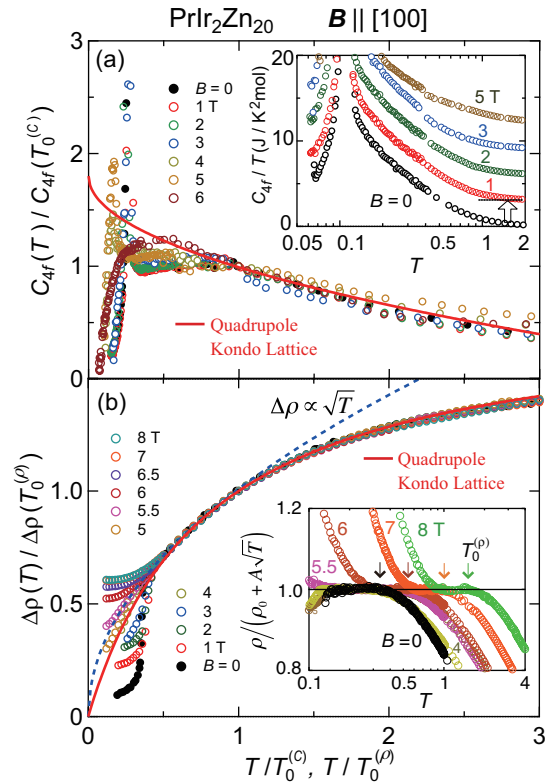}
\caption{(color online) Scaling plots of (a) the specific heat $C_{4f}$ and (b) the electrical resistivity ${\Delta}{\rho}$ of {\prirzn} in the magnetic fields \textbf{\textit B}${||}$[100].
In the temperature region 0.8$~{<}~{T}/{T}_{0}^{(C)},~{T}/{T}_{0}^{(\rho)}~{<}~$3, $C_{4f}$ and ${\Delta}{\rho}$ in magnetic fields better follow the calculation by using a two-channel Anderson lattice model \cite{Tsuruta15} as shown with the (red) solid curves than the (blue) dotted curve calculated with the impurity quadrupole Kondo model\cite{Cox98}.
The inset of (a) shows the temperature dependence of $C_{4f}{/}{T}$ with vertical offsets.
The temperature variation of ${\rho}{/}({\rho}_0 {+} {A}\sqrt{T})$ is shown in the inset of (b). At the value of 1, ${\rho}(T)$ follows $\sqrt{T}$. 
The arrows show the characteristic temperature, ${T}_{0}^{({\rho})}$, where ${\rho}(T)$ deviates from ${\rho}_{0}{+}{A}\sqrt{T}$ relation. 
}
\label{Scale}
\end{figure}


The sharp peak at {\tq} $=$ 0.11 K for ${B}$ = 0 in Fig. \ref{HC} (a) is the manifestation of the AFQ order.\cite{Onimaru11,Ishii11,Higemoto12} 
As shown with the open triangles in the inset of Fig. \ref{HC} (b), upon applying magnetic field for ${B}{\ge}$3 T, the peak shifts to lower temperatures and disappears at ${B}{=}$4.5 T.
The behavior of the specific heat peak at {\tq} in the magnetic fields is much different from the previous report where the peak is split into two  for 1${\le}{B}{\le}$3 T.\cite{Onimaru11} 
This inconsistency results from an extrinsic effect of another grain contained in the previous sample whose crystalline axis is directed away from the $a$-axis of the dominant crystal.

On the other hand, as shown with the open triangles in Fig. \ref{HC} (a), $C_{4f}(T)$  in $B$ = 6 T shows a broad peak at 0.6 K, which shifts to 2 K with increasing $B$ up to 12 T.
The height and width can be explained by taking account of the Zeeman splitting of the ground state doublet.
Because the split singlets lose quadrupolar degrees of freedom, no phase transition occurs in ${B}$ ${>}$ 6 T.



We pay attention to another broad peak in $C_{4f}$($B$ = 0) at around 0.4 K in Fig. \ref{HC}(a). When $C_{4f}{/}{T}$ is plotted vs ln$T$ in the inset of Fig. \ref{Scale}(a), we find the $-$ln$T$ dependence between 0.2 and 0.8 K. 
Since this $-$ln$T$ dependence of $C_{4f}{/}{T}$ emerges below 2 K, it arises from the degrees of freedom of the ${\Gamma}_3$ ground state.
Here, we define a characteristic temperature, ${T}_{0}^{(C)}$, as the temperature where $S_{4f}$ reaches 
$\frac{3}{4}$$R$ln2. 
This definition follows the Cox's definition of the Kondo temperature $T_{\rm K}$ for the impurity quadrupole Kondo system.\cite{Cox98}
As shown in Fig. \ref{HC}(b), a defined ${T}_{0}^{(C)}$ remains at around 0.4 K up to ${B}{=}$3 T and increases for ${B}{>}$3 T (see Fig. \ref{Diagram} (a)).
We plot in Fig. \ref{Scale} (a) the values of $C_{4f}(T){/}C_{4f}(T_{0})$ at various fields, which are well scaled with respect to ${T}{/}{T}_{0}^{(C)}$ in the range of ${T}{/}{T}_{0}^{(C)}{>}$0.8.
This scaling behavior will be discussed in detail by combining the NFL behavior of the electrical resistivity.


\subsection{Electrical resistivity}

\begin{figure}[b]
\includegraphics[scale=0.5]{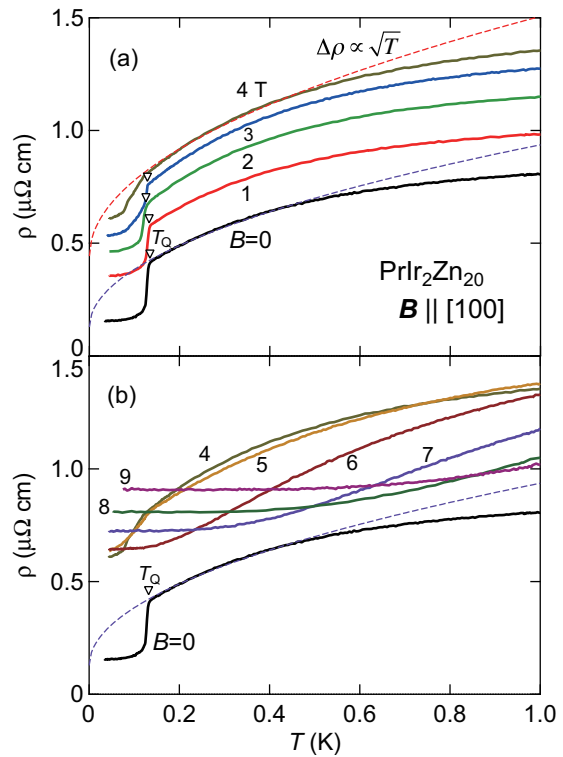}
\caption{(Color online) Temperature dependence of the electrical resistivity of {\prirzn} in magnetic fields of (a) 0${\le}{B}{\le}$4 T and (b) ${B}{\ge}$4 T applied along the [100] direction.
The open triangles indicate the AFQ ordering temperature, {\tq}.
The dashed lines are fits to the data using the relation of ${\Delta}{\rho}$ ${\propto}$ $\sqrt{T}$.
It is noted that the data are plotted without offset.}
\label{rho01}
\end{figure}

The temperature dependence of the electrical resistivity ${\rho}(T)$ in magnetic fields between 0 and 9 T applied along the [100] direction are shown in Fig. \ref{rho01}.
It is noted that all the data are plotted without offset.
In ${B}{\le}$4 T, ${\rho}(T)$ shows upward convex curvature at {\tq} $<$ $T$ $<$ 1.0 K, that is the NFL behavior.
On cooling below 0.5 K, ${\rho}(T)$ is likely to follow the $\sqrt{T}$ variation as shown with the dashed lines.
The absolute value of ${\rho}(T)$ for ${B}{\le}$4 T is increased with increasing the magnetic field.
On the other hand, above 6 T, ${\rho}(T)$ shows downward curvature.
The residual resistivity is increased with increasing the magnetic field.

\begin{figure}[t]
\includegraphics[scale=0.5]{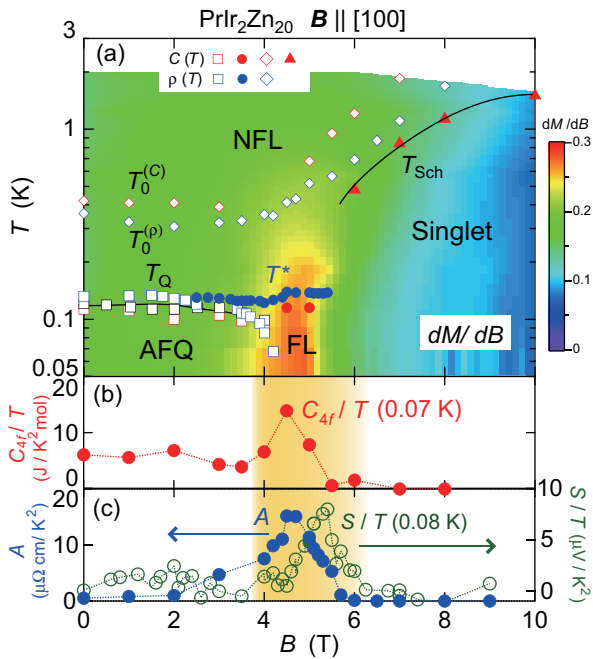}
\caption{(color online) (a) Magnetic-field vs temperature phase diagram of {\prirzn} for \textbf{\textit B}$||$[100]. 
The contour plot is made from the derivative of the magnetization with respect to the magnetic field, d${M}{/}$d${B}$, as shown in Fig. \ref{MH} (b).
The NFL state passes through the region denoted by ${T}^{*}$ to the field-induced Fermi liquid ground state, FL.
${T}_{0}^{({C})}$ and ${T}_{0}^{({\rho})}$ are the characteristic temperatures for the NFL behavior determined by the measurements of the specific heat and the electrical resistivity, respectively.
$T_{\rm Sch}$ is the temperature where a Schottky-type peak appears due to the splitting of the ground state doublet.
(b) Magnetic field variation of the specific heat divided by temperature, $C_{4f}{/}{T}$ at around 0.07 K. (c) Coefficient $A$ of the electrical resistivity, ${\rho}{=}{\rho}_{0}{+}{A}{T}^{2}$ (left-hand scale) and the Seebeck coefficient divided by temperature, ${S}{/}{T}$, at 0.08 K (right-hand scale)\cite{Ikeura14}. }
\label{Diagram}
\end{figure}

We pay our attention again to the NFL behavior of the upward convex curve at {\tq} $<$ $T$ $<$ 1.0 K.
Here, the data of ${\rho}(T)$ are replotted in the inset of Fig. \ref{Scale} (b) as ${\rho}{/}({\rho}_0 {+} A \sqrt{T})$ vs $T$, where ${\rho}_0$ is the residual resistivity.
In the $T$ region where ${\rho}{/}({\rho}_0 {+} A \sqrt{T})$ stays at 1.0, ${\Delta}{\rho}{=}{\rho}(T){-}{\rho}_{0}$ follows $\sqrt{T}$. 
The arrows denote the characteristic temperature, ${T}_{0}^{({\rho})}$, where ${\rho}(T)$ starts deviating from the $\sqrt{T}$ dependence on heating. 
We represent the data of ${\Delta}{\rho}(T){/}{\Delta}{\rho}(T_{0})$ vs ${T}{/}{T}_{0}^{(\rho)}$ in Fig. \ref{Scale} (b), where ${\rho}(T)$ follows the scaling well in the temperature range of 0.5 $<$ ${T}{/}{T}_{0}^{(\rho)}$ $\le$ 3.
As shown in the inset of Fig. \ref{Scale} (b), ${T}_{0}^{({\rho})}$ stays at around 0.35 K up to ${B}{=}$4 T, and increases significantly once ${B}$ exceeds 4 T (see the (red) opened diamonds in Fig. \ref{Diagram} (a)). 
The field dependence of ${T}_{0}^{({\rho})}$ coincides with that of ${T}_{0}^{(C)}$ described in the previous subsection, suggesting the same origin of the both ${T}_{0}^{({\rho})}$ and ${T}_{0}^{(C)}$.\cite{Stewart01}

\subsection{Non-Fermi liquid behavior for $T>T_{\rm Q}$}

Let us discuss possible mechanisms for the observed NFL behaviors of $C_{4f}(T)$ and ${\Delta}{\rho}(T)$ in the magnetic fields well scaled with respect to ${T}/{T}_{0}^{(C)}$ and ${T}/{T}_{0}^{(\rho)}$, respectively.
One is the contribution of a rattling phonon, which gives rise to ${\Delta}{\rho}{\propto}\sqrt{T}$.\cite{Dahm07}
There is an optical phonon excitation at around 7 meV (80 K), which is attributed to the low-energy vibration of the Zn atom.\cite{Hasegawa12,Wakiya15, Wakiya16} 
However, as this phonon excitation probably ceases at ${T}{<}$1 K, interaction of the optical phonon mode with the conduction electrons must be too weak to modify the low-temperature electronic transport.

The second is the impurity quadrupole Kondo effect predicted by Cox {\it et al.}, which leads to ${\Delta}{\rho}(T){\propto}{1}{+}{A}\sqrt{T}$ and ${C}{/}{T}{\propto}{-}{\rm ln}{T}$ as mentioned in the introduction.\cite{Cox98} 
The $\sqrt{T}$ dependence of ${\rho}(T)$ may be applicable to the present data.
However, in the present case of {\prirzn}, the quadrupole moments are not included as impurities but placed periodically, therefore, formation of the quadrupole Kondo lattice is a promising candidate bringing about the NFL state.\cite{Jarrell96}
The theoretical analyses on the temperature variations of ${\rho}$ and ${C}$ with the two-channel Anderson lattice model have led the following relations;
\begin{eqnarray}
{\Delta}{\rho}(T) &=& \frac{a_{1}}{1+{a_{2}} ({T}_{0}^{({\rho})}{/}{T}) }\\
C(T) &=& b_{1} \left(1{-}b_{2}\sqrt{\frac{T}{{T}_{0}^{(C)}}}\right), 
\end{eqnarray}
where $a_{i}$ and $b_{i}$ (${i}{=}$1 and 2) are parameters.\cite{Tsuruta15}
The calculations are shown by the (red) solid curves in Fig. \ref{Scale} (a) and (b).
The curves are in better agreement with our experimental data for wider temperature range than those calculated by the impurity quadrupole Kondo model as shown with the (blue) dotted line.

In the present case, the NFL behavior does not persist down to the low-temperature limit. This is certain because this NFL behavior does not arise from the conventional quantum critical point but from the quadrupole Kondo effect as described in the introduction. The quadrupoles are over-screened by the conduction electrons, leading to the unstable electronic state with the residual entropy of (0.5)$R$ln2 at zero temperature in principle. In real systems, however, the residual entropy should go to zero at  $T$ = 0 by a mechanism following the third law of thermodynamics. In fact, in the magnetic fields $B$ $\le$ 4 T, the residual entropy is released by the intersite quadrupole interaction, leading to occurrence of the quadrupole order. The NFL behavior certainly manifests itself in the limited temperature range, where the magnetic entropy goes down from $R$ln2 to (0.5)$R$ln2. In zero magnetic field, where the NFL behaviors of the electrical resistivity and the specific heat appear in the temperature range from 1.5 K to 0.2 K.

Taking these experimental results and the analyses on them, this is the first observation of the NFL state in the nonmagnetic quadrupole-active system, whose origin is intrinsically different from that observed in the vicinity of the conventional quantum critical point. 
The scaling plots of $C(T)$ and ${\Delta}{\rho}(T)$ using the characteristic temperatures ${T}{/}{T}_{0}^{(C)}$ and ${T}{/}{T}_{0}^{(\rho)}$, respectively, as shown in Fig. \ref{Scale}(a) and (b), imply the possible formation of the quadrupole Kondo lattice resulting from hybridization between the quadrupoles and the conduction electrons with an energy scale of $k_{\rm B}{T}_{0}$.

\begin{figure}[t]
\includegraphics[scale=0.5]{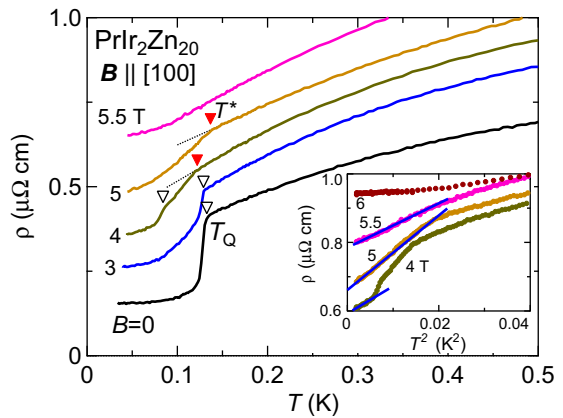}
\caption{
(color online) Temperature dependence of the electrical resistivity ${\rho}(T)$ of {\prirzn} in the magnetic field of 0${\le}{B}{\le}$5.5 T applied along the [100] direction. 
The data in the magnetic fields are offset for clarity.
The open and solid triangles indicate the AFQ order temperature {\tq} and the crossover temperature $T^{*}$, respectively. 
The inset is the plot of ${\rho}$ vs $T^2$ in $4~{\le}~B{\le}~6$ T with vertical offset.
}
\label{rho}
\end{figure}

\subsection{Magnetic-field induced Fermi liquid state}

Looking close to the data of ${\rho}(T)$ at ${B}$ = 4 T as shown with the solid triangles in Fig. \ref{rho}, we find another knee at ${T}^{*}$ = 0.12 K above {\tq} = 0.08 K.
In contrast to the drop of {\tq} at ${B}$ ${>}$ 3 T, ${T}^{*}$ stays at around 0.13 K for ${B}$ ${=}$ 4 and 5 T.
As shown in the inset of Fig. \ref{HC} (b), this ${T}^{*}$ coincides with the broad peak of $C_{4f}(T)$ at ${T}^{*}$ = 0.12 K in ${B}$ = 4.5 and 5 T.
The field dependences of  ${T}^{*}$ observed in the ${\rho}(T)$ and ${C}_{4f}(T)$ measurements are plotted with the blue and red closed circles, respectively, in the ${B}{-}{T}$ phase diagram in Fig. \ref{Diagram} (a).
The ${T}^{*}$ line can be a boundary of two states, meaning a phase transition or a crossover. 
In the present case, the crossover is likely to occur at ${T}^{*}$, because the peak of ${C}_{4f}(T)$ becomes very broad despite the rather robust ${T}^{*}$ against the magnetic field.
On cooling below ${T}^{*}$, ${\rho}(T)$ follows ${\rho}_{0}{+}{A}{T}^2$ at low temperature for 4${\le}{B}{<}$6 T, as shown in the inset of Fig. \ref{rho}.
The $A$ coefficient as a function of $B$ is plotted in Fig. \ref{Diagram} (c).
It is strongly enhanced at around 4.5 T, where $C_{4f}{/}{T}$ at 0.07 K is also peaked.
Moreover, there is a peak at 5.5 T in the Seebeck coefficient divided by temperature, ${S}{/}{T}$, at 0.08 K.\cite{Ikeura14}
The coincidence of these peaks at around 5 T suggests a peculiar feature of the heavy fermion state on cooling through ${T}^{*}$ in a narrow range of $B$, 3.5${<}{B}{<}$6 T.

\begin{figure}[t]
\includegraphics[scale=0.5]{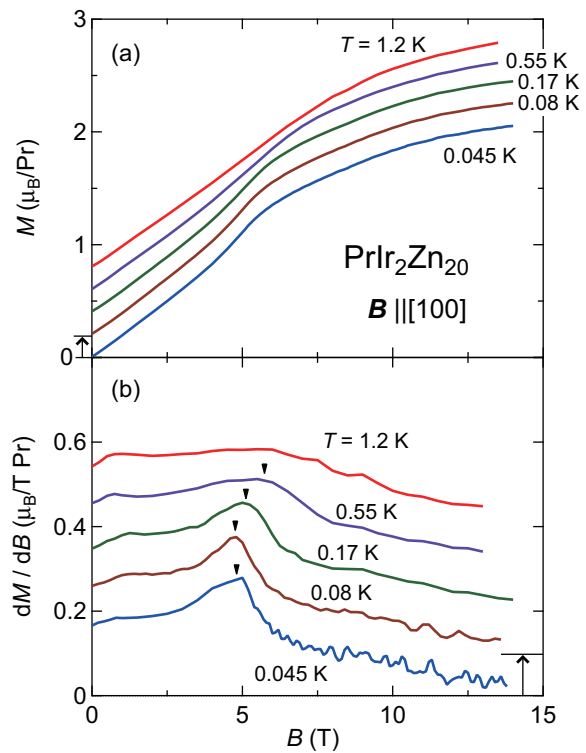}
\caption{(Color online) (a) Isothermal magnetization and (b) d${M}{/}$d${B}$ of {\prirzn} at temperatures between 0.045 and 1.2 K in the magnetic field ${B}{\le}$14 T applied along the [100] direction. The data are vertically offset for clarity. 
The solid and dotted lines indicate the data for ascending the magnetic fields.}
\label{MH}
\end{figure}

Fig. \ref{MH} (a) and (b) show the isothermal magnetization and the derivative of the magnetization with respect to the magnetic field, d${M}{/}$d${B}$, at ${T}{=}$0.045, 0.08, 0.17, 0.55, and 1.2 K in the magnetic field of ${B}{\le}$14 T applied along the [100] direction, respectively.
The isothermal magnetization shows metamagnetic behavior at around 5 T, where
d${M}{/}$d${B}$ shows a peak. 
The peak shifts slightly to higher fields and becomes broader with increasing temperatures, which is likely to be connected to the Schottky anomaly due to the splitting of the $\Gamma_{3}$ doublet in the magnetic field for ${B}{>}$6 T.

The contour plot in Fig. \ref{Diagram} (a) indicates the derivative of the magnetization with respect to the magnetic field, d${M}{/}$d${B}$.
Note that the d${M}{/}$d${B}$ is largely enhanced only in the narrow range of $B$, 3.5${<}{B}{<}$5.5 T.
Furthermore, the elastic modulus of the $C_{11}$ mode, which is a part of the ${\Gamma}_3$-type $(C_{11}{-}C_{12}){/}2$ mode, shows strong softening in the magnetic field at around ${B}{=}$5 T.\cite{Ishii11}
All these observations suggest development of an exotic Fermi liquid state on cooling through ${T}^{*}$ in a range of $B$, 3.5${<}{B}{<}$5.5 T.

There are some possible mechanisms to form the exotic Fermi liquid state at ${T}{<}{T}^{*}$.
One is a crossover from the NFL state to a FL state expected in the two-channel Kondo model taking in magnetic field effect.\cite{Yotsuhashi02}
Perhaps, emergence of a FL state accompanied with a free magnetic spin in the vicinity of the NFL state has been pointed out by taking into account possible competition between the magnetic and quadrupole Kondo effects on the basis of an extended two-channel Kondo model.\cite{Kusunose14}
Another is a cross-over due to the composite electronic order \cite{Hoshino11,Hoshino13} or the Hastatic order \cite{Hastatic} in the quadrupole Kondo lattice as were mentioned above.
In any cases, it is noted that the hybridization between the quadrupoles and the conduction electrons gives rise to a new type of the electronic ground state.
It is highly desirable to detect the order parameter by microscopic techniques.

Indeed, we expected that the quadrupole induced quantum critical behavior in $C_{4f}(T)$ would appear when the AFQ order is fully suppressed. 
However, as shown in Fig. \ref{Diagram} (b), $C_{4f}{/}{T}$ does not diverse in the vicinity of 4.5 T on cooling to 0.07 K at all. It seems that the Doniach picture is not valid for this pure quadrupole system.
Although it is still an open question why the quantum criticality was not detected in {\prirzn},
a peculiar feature of the quadrupole must give rise to the unique electronic phenomena.

\section{Summary}

We report the low-temperature transport, thermodynamic and magnetic properties on a cubic system {\prirzn} in which the non-Kramers doublet ground state has purely the quadrupolar degrees of freedom.
In the moderately wide temperature range at ${T}{>}${\tq}, non-Fermi liquid behaviors were clearly observed in the electrical resistivity ${\rho}$ and the specific heat $C_{4f}$.
Both ${\rho}$ and $C_{4f}$ in magnetic fields ${B}$ ${<}$ 6 T applied along the [100] direction can be well scaled with a characteristic temperature $T_{0}$, suggesting the formation of the quadrupole Kondo lattice due to the hybridization between the quadrupoles and the conduction electrons.
It indicates that the NFL behavior observed in the present system is intrinsically different from that observed in the vicinity of the conventional quantum critical point.
Furthermore, $\rho$ and $C_{4f}$ exhibit anomalies at ${T}^{*}$$=$0.13 K in the vicinity of 5 T, where the coefficient $A$ for ${\rho}{=}{\rho}_{0}{+}{A}{T}^{2}$, ${C}_{4f} / {T}$, and ${S} {/} {T}$, have significant enhancement as a function of $B$. 
The concomitant increase in d${M}{/}$d$B$ indicates formation of a magnetic-field induced Fermi liquid ground state to remove the residual entropy in the quadrupole Kondo lattice.
These observations imply that the Doniach picture relevant to the spin Kondo systems should not be valid for this purely quadrupole-active system, possibly leading us beyond the Doniach picture.

\section*{Acknowledgements}

The authors would like to thank A. Tsuruta, K. Miyake, H. Kusunose, S. Hoshino, J. Otsuki, Y. Kuramoto, K. Uenishi, Y. Yamane, I. Ishii, T. Suzuki and K. Iwasa for helpful discussions.
We also thank Y. Shibata for the electron-probe microanalysis performed at N-BARD, Hiroshima University. 
The magnetization measurements with MPMS and specific heat measurements with PPMS were carried out at N-BARD, Hiroshima University. This work was financially supported by JSPS KAKENHI Grant Numbers 21102516, 23102718, 26707017, and 15H05884, 15H05886 (J-Physics), and by The Mazda Foundation Research Grant, Japan and Hiroshima University Fujii Research Fund.


\end{document}